\documentclass[12pt,a4paper]{article}
\usepackage{graphicx}
\usepackage{amssymb}
\usepackage{amsmath}
\usepackage{a4}
\usepackage{setspace}
\usepackage{color}

\title{Theory of Light Sail Acceleration by Intense Lasers: an Overview}
\author{Andrea Macchi\\
\emph{National Institute of Optics, National Research Council (CNR/INO)},\\\emph{research unit ``Adriano Gozzini''}, \and %\\ 
\emph{Department of Physics ``Enrico Fermi'', University of Pisa},\\
\emph{Largo Bruno Pontecorvo 3, I-56127 Pisa, Italy}\\
E-mail: {\tt andrea.macchi@ino.it}, Web: {\tt www.df.unipi.it/$\sim$macchi}
}

\begin{document}

\maketitle

\begin{abstract}
A short overview of the theory of acceleration of thin foils driven by the radiation pressure of superintense lasers is presented. A simple criterion for radiation pressure dominance at intensities around $5 \times 10^{20}~\mbox{W cm}^{-2}$ is given, and the possibility for fast energy gain in the relativistic regime is discussed. 
\end{abstract}

\doublespacing

\section{Introduction}

It has been known since the discovery of Maxwell's equations that light, i.e. electromagnetic (EM) radiation exerts a pressure on a reflecting object, and thus may accelerate it. In 1925, F.~Zander suggested to exploit the radiation pressure of the Sun for space travel using light sails, i.e. mirrors of large area and small thickness. 

The scattering of an EM wave by a particle also leads to momentum absorption and acceleration. In 1957, V.~I.~Veksler \cite{vekslerSJAE57} suggested that Thomson scattering by a small cluster containing $N$ electrons may accelerate the cluster to high velocities. The fundamental point of Veksler's proposal was that the radiation force on the cluster scaled as $N^2$, providing an example of his new principle of coherent acceleration, i.e. the use of collective effects to accelerate large amounts of particles to high energies. 

\begin{figure}[t!]
\begin{center}
\includegraphics[width=0.25\textwidth]{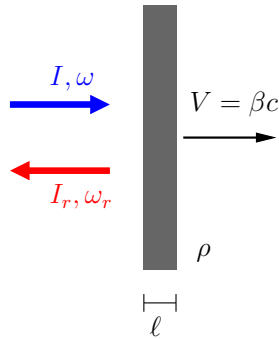}
\end{center}
\caption{The light sail concept. The sail is modeled as a perfect mirror of surface density $\sigma=\rho\ell$ with $\rho$ the mass density and $\ell$ the thickness. The sail is pushed by a plane wave of intensity $I$ and frequency $\omega$. Notice that the equation of motion for the sail (\ref{eq:LS}) and the expression for the mechanical efficiency may be simply obtained by considering the Doppler shift of the reflected radiation [$\omega_r=\omega(1-\beta)/(1+\beta)$] and the conservation of the ``number of photons'', see e.g. Ref.\cite{macchi-book-ionacceleration}.  }
\label{fig:LSscheme}
\end{figure}

After the invention of the laser, R.~L.~Forward in 1962 \cite{forwardJS84,gilster-book-forward} and G.~Marx in 1966 \cite{marxN66} proposed to use an Earth-based laser system to accelerate a rocket up to relativistic velocities. Marx's paper included a relativistic analysis of the motion of a sail, i.e. a plane perfect mirror, accelerated by radiation pressure, based on the equations
\begin{equation}
\frac{d(\gamma V)}{dt}=\frac{2}{\sigma_0 c}I(t-X/c)\frac{1-V/c}{1+V/c}
\; , \qquad
\frac{dX}{dt}=V \; ,
\label{eq:LS}
\end{equation}
where $I=I(t)$ is the intensity of the laser pulse, $\sigma_0$ is the surface mass density of the sail, and $\gamma=(1-V^2/c^2)^{-1/2}$. The concept is sketched in Fig.\ref{fig:LSscheme}. The most interesting result (but also the subject of a long-lasting controversy \cite{simmonsAJP92}) was the expression for the mechanical efficiency 
$\eta=2\beta/(1+\beta)$ (with $\beta=V/c$), that reaches 100\% in the relativistic limit $\beta \to 1$. Eqs.(\ref{eq:LS}), hereafter refereed to as the light sail (LS) equations, have the same form as for the motion of the Thomson scattering particle \cite{landau-mirror}, evidencing the connection with Veksler's proposed mechanism.

In 2004, using particle-in-cell (PIC) simulations of the acceleration of a thin plasma foil by a laser pulse with intensity $I>10^{23}~\mbox{W cm}^{-2}$ Esirkepov et al. \cite{esirkepovPRL04} showed that the motion of the foil was also well fitted by the above mentioned equation, giving evidence that the foil was driven from radiation pressure. The scaling of the LS equation to foreseeable laser and target parameters showed the possibility to reach relativistic velocity of the foil, corresponding to an energy per nucleon above the GeV barrier. The coherent motion of the foil also implied an inherent mono-energetic spectrum, which would be crucial for most applications. Such features have then stimulated a strong interest in LS acceleration.

In this paper we give a brief overview of the research on LS acceleration in the past decade, mostly focusing on theoretical aspects and open issues. A simple criterion for radiation pressure dominance at intensities around $5 \times 10^{20}~\mbox{W cm}^{-2}$ is given, and the possibility for fast energy gain in the relativistic regime is pointed out. A more comprehensive presentation of experimental and simulation results may be found in recent review papers on laser-driven ion acceleration \cite{daidoRPP12,macchiRMP13,macchiPPCF13,fernandezNF14}.

\section{One-dimensional dynamics}

For an arbitrary pulse profile $I(t)$, the final value of $\gamma$ is obtained from Eq.(\ref{eq:LS}) as
\begin{equation}
\gamma_{\infty}\equiv\gamma(t={\infty})=1+\frac{{\cal F}^2}{2({\cal F}+1)} \; , \qquad {\cal F}=\frac{2}{\sigma c^2}\int_0^{\infty}I(t')dt' \; . 
\label{eq:gammaLS}
\end{equation}
For a flat-top intensity profile, i.e. a constant value of $I$ between $t=0$ and $\tau_L$, Eq.(\ref{eq:LS}) can be solved exactly. Here we just give the limiting cases of $\beta\ll 1$ and $\beta\to 1$, for which the integration is straightforward (notice that for $\beta \simeq 1$ then $(1+\beta)/(1-\beta) \simeq 4\gamma^2$):
\begin{equation}
\gamma(t)=\left\{
\begin{array}{ll}
1+[1-\exp(-2\Omega t)]^2/8 & (\Omega t\ll 1) \\
(3\Omega t/4)^{1/3} & (\Omega t\gg 1) 
\end{array}
\right. \; ,
\label{eq:gammaLSt}
\end{equation}
where $\Omega=2I/\sigma_0 c^2$. Eqs.(\ref{eq:gammaLS}) and (\ref{eq:gammaLSt}) may be used to obtain the acceleration time and length in the laboratory for a given value of the final energy per nucleon ${\cal E}_{\mbox{\tiny max}}=m_pc^2(\gamma_{\infty}-1)$ (notice that it would be incorrect to plug the pulse duration $\tau_L$ in Eq.(\ref{eq:gammaLSt}) to obtain ${\cal E}_{\mbox{\tiny max}}$). 

It is evident that the energy gain is quite fast for $\beta\ll 1$ but becomes much slower in the relativistic regime as $\beta\to 1$. In a realistic multi-dimensional scenario this is a possible issue because of laser pulse diffraction on distances larger than the Rayleigh length. Fortunately, as discussed below the energy gain may be faster in 3D geometry thanks to the target rarefaction.

Obviously, the lighter the sail the higher the energy for a given laser pulse. However, if the foil target is too thin then it becomes transparent to the laser pulse and the radiation pressure boost drops down. Based on the simple model of a delta-like foil and purely transverse electron motion \cite{vshivkovPoP98,macchi-book-SIT}, the threshold for transparency due to relativistic effects is given by 
\begin{equation}
a_0 \simeq \zeta \; ,
\label{eq:a0zeta}
\end{equation} 
where $a_0=(I/m_e n_cc^3)^{1/2}$, $\zeta=\pi\sigma_0/(Zm_in_c\lambda)$, $n_c=\pi m_ec^2/(e^2\lambda^2)=\pi/(r_c\lambda^2)$ is the cut-off density and $\lambda$ is the laser wavelength. Despite the very simplified underlying model, Eq.(\ref{eq:a0zeta}) describes fairly well the onset of transparency and the breakdown of LS acceleration observed in 1D simulations \cite{macchiPRL09}. Actually Eq.(\ref{eq:a0zeta}) may be considered as slightly pessimistic because as the foil moves the reflectivity increases due to the decrease of the pulse frequency in the moving frame \cite{macchiNJP10}. The situation is more complex for finite width pulses in  multidimensional geometry because the transverse expansion of the foil leads to a decrease of the surface density $\sigma$ in time.

\begin{figure}[t!]
\begin{center}
\includegraphics[width=0.8\textwidth]{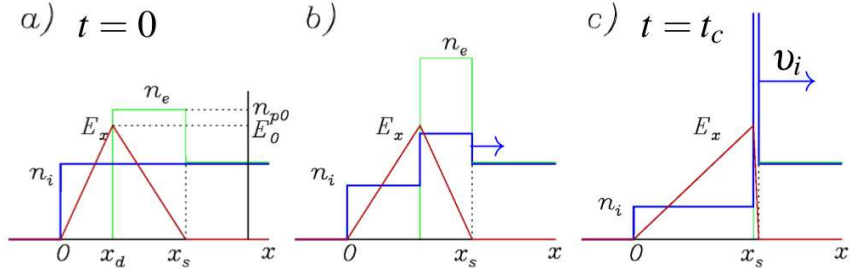}
\end{center}
\caption{Cartoon sketching the first stage of ion acceleration driven by radiation pressure \cite{macchiPRL05}. The densities of ions ($n_i$) and electrons ($n_e$) are approximated by step-like functions. Ions initially in the $x_d<x<x_s$ layer are accelerated by the charge separation field $E_x$ up to the velocity $\upsilon_i$ at the time $t=t_c$.}
\label{fig:RPAcartoon}
\end{figure}

The above modeling considers the sail as a neutral, rigid body with electrons comoving with ions. Indeed, charge separation effects are crucial in the ``inner'' dynamics of LS acceleration. The cartoon in Fig.\ref{fig:RPAcartoon} describes the initial stages of ion acceleration. Electrons are pushed into the target by the secular ponderomotive force per unit volume ${\bf f}_p=\left\langle{\bf J}\times{\bf B} \right\rangle$, where the brackets denote a  cycle average.  The ponderomotive force sweeps and piles up the electrons creating a charge depletion layer until ${f}_p$ is exactly balanced by an electrostatic field $E_x$ (this corresponds to the balance between $P_{\mbox{\tiny rad}}$ and the electrostatic pressure on ions, see e.g. Refs.\cite{macchiPRL09,macchiNJP10}). In turn, $E_x$ accelerates the ions as sketched in the cartoon of Fig.1. In a first stage, the ions in the layer where the EM field penetrates are accelerated up to a velocity $v_i$ within a time $t_c$, which are given by \cite{macchiPRL05}
\begin{equation}
\frac{\upsilon_i}{c} \simeq \left(\frac{I}{\rho c^3}\right)^{1/2}=\left(\frac{Zm_en_c}{Am_pn_e}\right)^{1/2}a_0 \; , \qquad t_c \simeq \frac{1}{\omega a_0}\left(\frac{Am_p}{Zm_e}\right)^{1/2} \; 
\label{eq:ionmotion}
\end{equation}
where $\rho$ is the mass density (for simplicity we assume non-relativistic motion; see \cite{robinsonPPCF09} for relativistically corrected expressions). At $t=t_c$ the accelerated ions have piled up at the position $x=x_s\simeq\upsilon_it_c$. If this position coincides with the rear surface of the foil, the acceleration cycle may be repeated, and eventually the sequence of acceleration stages converges to the motion described by Eq.(\ref{eq:LS}) \cite{grechNJP12}.

The correct balance of electrostatic and radiation pressure shows that only a fraction $F \simeq 1-a_0/\zeta$ of the ions is accelerated coherently as a sail, even if the motion of the latter is still described by Eq.(\ref{eq:LS}) with $\sigma_0$ including the \emph{total} mass of the foil \cite{macchiPRL09,macchiNJP10}. During the motion, as far as the geometry is 1D the electrostatic pressure on the sail only depends on the total charge behind the sail, while the radiation pressure decreases by a factor $(1-\beta)/(1+\beta)$. Thus, to maintain the pressure balance, additional ions are progressively trapped in the sail \cite{macchiNJP10,eliassonNJP09}. Applying the pressure balance as in Ref.\cite{macchiPRL09} with the velocity correction yields the final fraction of accelerated ions:
\begin{equation}
F \simeq 1-\frac{a_0}{\zeta}\left(\frac{1-\beta_{\infty}}{1+\beta_{\infty}}\right)^{1/2} \; ,
\end{equation}
where $\beta_{\infty}=2/[(1+{\cal F})^2+1)]$. This shows that all ions are eventually accelerated in the relativistic limit.

\section{Radiation pressure dominance}

Since a thin plasma foil is not a perfect mirror, it is not trivial that irradiation by intense light should result in LS acceleration. In most of accessible laser-plasma interaction conditions, strong heating of electron occurs, and the resulting kinetic pressure exceeds radiation pressure; in such situation, the plasma foil expands and the resulting ion energy spectrum is very different from the LS case. The situation is somewhat reminiscent of the Crookes radiometer or light mill where the vanes are white (reflecting) on one side and black (absorbing) on the other side: the mill rotates in the direction \emph{opposite} to what would be expected from radiation pressure being higher on the white side than on the black side, because the effects of heating and thermal pressure dominate. 

To find the conditions in which the radiation pressure $P_{\mbox{\tiny rad}}$ will dominate the acceleration let us briefly recall the heating dynamics of electrons. At normal incidence, electrons are driven in the direction perpendicular to the target surface by the ${\bf v}\times{\bf B}$ force which for linear polarization (LP) has an oscillating term at $2\omega$ (where $\omega$ is the laser frequency) in addition to the secular ponderomotive force. Heating of electrons occurs via oscillations across the laser-plasma interface driven by the oscillating term, which vanishes for circular polarization (CP) \cite{macchiPRL05}. The use of CP pulses has then been proposed by several authors \cite{zhangPoP07,klimoPRSTAB08,robinsonNJP08} to obtain an efficient LS regime at ``any'' intensity. Detailed 3D simulations in the relativistic regime \cite{tamburiniPRE12} also showed that for CP pulses higher energies and better collimation of the ion beam are obtained with respect to LP pulses. Experiments performed so far, however, have shown a limited impact of the use of CP \cite{henigPRL09,dollarPRL12,karPRL12,aurandNJP13} and non-LS effects such as species separation in the spectrum  \cite{karPRL12,aurandNJP13,steinkePRSTAB13} (in the ideal LS regime, all species move at the same velocity, thus the energy per nucleon is independent on the mass number). These data suggest that tight focusing of the laser pulse and, possibly, imperfect conversion to CP may prevent efficient LS operation, at least in the intensity regime investigated so far, i.e. $I \simeq (2\times 10^{18}\div 2 \times 10^{21})~\mbox{W cm}^{-2}$.

In view of future experiments at higher intensities and of possible technical difficulties for producing ultraintense CP pulses, it appears important to discuss possible conditions for radiation pressure dominance also for LP, when electron heating is important. 
Heuristically, the transfer of energy to ions via $P_{\mbox{\tiny rad}}$ can be efficient if it is ``faster'' than the heating of electrons, which occurs on a laser halfcycle. Esirkepov et al. \cite{esirkepovPRL04} suggested that ions should become promptly relativistic, i.e. reaching a velocity close to $c$ within one cycle, so that they would ``stick'' to electrons. To estimate the corresponding laser intensity for such regime, let us assume $\upsilon_i \simeq c/2$ in Eq.(\ref{eq:ionmotion}): this gives
\begin{equation}
a_0 \simeq 30\left(\frac{n_e}{n_c}\right)^{1/2}
\; ,
\end{equation}
which for $n_e/n_c \simeq 100$ gives $I\lambda^2> 10^{23}~\mbox{W cm}^{-2}\mu\mbox{m}^2$, that is the typical intensity of the simulations in Ref.\cite{esirkepovPRL04}. These values are not presently available although they may be reached in the laboratory within the next decade. 

Here we propose a different condition which leads to a more accessible intensity threshold. The above defined ion acceleration time $t_c$ may be taken as the relevant temporal scale for energy transfer to ions. For electrons, acceleration occurs on a laser halfcycle being driven by the oscillating force at $2\omega$. Thus we suggest $t_c<\pi/\omega$ as the condition for energy transfer to ions being more efficient than to electrons. This leads to the threshold for the laser amplitude
\begin{equation}
a_0 >\frac{1}{\pi}\left(\frac{Am_p}{Zm_e}\right)^{1/2} \simeq 19 \; ,
\end{equation}
which is equivalent to $I\lambda^2>5 \times 10^{20}~\mbox{W cm}^{-2}\mu\mbox{m}^2$, independently of the plasma density. This estimate is in qualitative agreement with LS signatures being observed in current experiments at similar intensities \cite{karPRL12}. A slightly greater intensity threshold of $10^{21}~\mbox{W cm}^{-2}\mu\mbox{m}^2$ has been suggested by Qiao et al \cite{qiaoPRL12} on a different basis, i.e. by comparing the ion energy gain due to the radiation pressure push with that in the fast electron sheath. 

\section{Fast gain regimes: ``unlimited'' acceleration}

In a realistic situation the laser pulse has a finite width and drives a cocoon deformation and transverse expansion of the target. This unavoidable effect may lead to early breakthrough of the laser pulse and termination of LS stage, thus the use of smooth transverse profile to keep a nearly plane geometry was suggested by several simulation studies. In contrast, Bulanov et al. \cite{bulanovPRL10} suggested that the decrease of target density due to transverse expansion may lead, in proper conditions, to acceleration up to higher energies than in the planar case, at the expense of the number of accelerated ions. This has been named as the ``unlimited'' acceleration regime. 

In the following we give a brief and simplified account of the detailed theory developed by Bulanov et al. \cite{bulanovPoP10}. The basic modification of Eq.(\ref{eq:LS}) for the longitudinal motion of the sail is that the surface density now depends on time due to the transverse expansion, 
\begin{equation}
\sigma=\sigma(t)=\frac{\sigma(0)}{\Lambda^{D-1}(t)} \; ,
\end{equation}
where $\Lambda(t)$ describes the dilatation of the transverse position of a fluid element of the sail, i.e. $r_{\perp}(t)=\Lambda(t)r_{\perp}(0)$, and $D$ is the dimensionality of the system; $D=1$ corresponds to planar geometry (constant $\sigma$), $D=2$ to two-dimensional Cartesian geometry, and $D=3$ to three-dimensional geometry with cylindrical symmetry. Now, it is assumed that for a given element the motion is ballistic after an initial kick by the laser pulse delivering a transverse momentum $p_{\perp}=m_i\varpi_0r_{\perp}(0)$, i.e. proportional to the initial position. This relation might be justified by observing that such kick comes from the transverse ponderomotive force, which is proportional to the gradient of the intensity and thus would be a linear function of position for a parabolic profile. It is further assumed that $p_{\perp} \ll p_{\parallel}$, with $p_{\parallel}$ the longitudinal momentum. The transverse velocity thus decreases as a result of the increasing longitudinal momentum. This leads to the equation for $\Lambda$
\begin{equation}
\frac{d\Lambda}{dt}=\frac{\dot{r}_{\perp}(t)}{r_{\perp}(0)}
=\frac{\varpi_0}{\gamma(t)} \; , \qquad
\gamma(t) \simeq (p^2_{\parallel}+m_i^2c^2)^{1/2} \; ,
\label{eq:unlimit1}
\end{equation}
with the coupled equation for $p_{\parallel}=\gamma\beta_{\parallel}$:
\begin{equation}
\frac{d(\gamma \beta_{\parallel})}{dt}=\frac{2I}{\sigma_0 c^2}\Lambda^{D-1}(t)\frac{1-\beta_{\parallel}}{1+\beta_{\parallel}} \; .
\label{eq:unlimit2}
\end{equation}
For further simplification we consider the asymptotic, ultrarelativistic limit in which $\beta_{\parallel}\to 1$ and $({1-\beta_{\parallel}})/({1+\beta_{\parallel}}) \simeq (2\gamma)^{-2}$. In this limit we find a solution
\begin{equation}
\gamma=\left(\frac{t}{\tau_k}\right)^k \; , \qquad k=\frac{D}{D+2} \; ,
\label{eq:LSunlimitedscalings}
\end{equation}
with the time constants given for $D=1,2,3$ by
\begin{equation}
\tau_{1/3}=\left(\frac{3}{4\Omega}\right) \; , \qquad
\tau_{2/4}=\left(\frac{1}{\Omega\varpi_0}\right)^{1/2} \; , \qquad
\tau_{3/5}=\left(\frac{48}{125\Omega\varpi_0^2}\right)^{1/3} \; .
\end{equation}
Fast energy gain in this regime thus depends on the initial conditions via the parameter $\varpi_0$. Assuming the initial transverse kick to be of the same order as the longitudinal one, $\varpi_0\simeq\Omega$ may be assumed for a quick estimate. For a given temporal profile $I(t)$, the final energy and surface density along the axis may be obtained by integrating Eqs.(\ref{eq:unlimit1}-\ref{eq:unlimit2}) with respect to the proper time $t'=t-X/c$ with $dt'=(1-\beta_{\parallel})dt$. Bulanov et al. \cite{bulanovPoP10} also discuss ``optimal'' pulse profiles to maximize the acceleration; heuristically, since the ``unlimited'' mechanism is actually limited by the onset of transparency, one argues that the decrease of the density may be matched with the decrease of the pulse frequency in the sail frame to keep a constant reflectivity.

\section{Three dimensional simulations}

The above outlined theory shows that, differently from other acceleration mechanisms \cite{sgattoniPRE12,dhumieresPoP13} the energy gain may be higher in a realistic 3D geometry that with respect to a 1D plane case. Confirmation of this theory in numerical experiments thus requires fully 3D, large scale simulations which are feasible on the most powerful present-day parallel supercomputers.

\begin{figure}[t!]
\begin{center}
\includegraphics[width=0.8\textwidth]{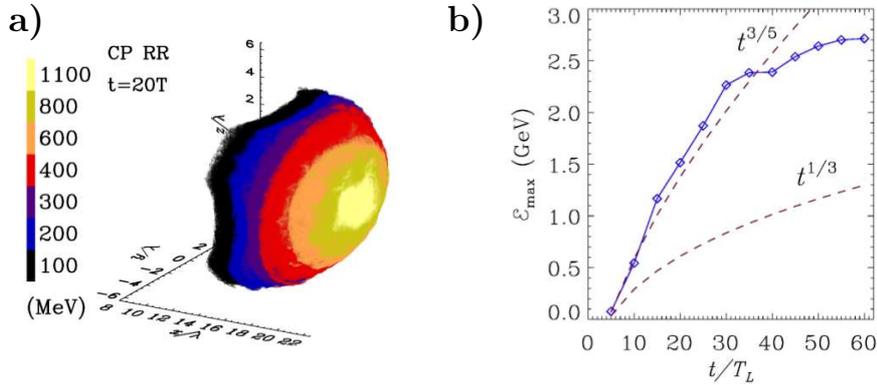}
\end{center}
\caption{Three dimensional particle-in-cell simulations of thin foil acceleration. Frame~a): space and energy distribution of ions \cite{tamburiniPRE12} (reproduced by permission of APS) at $t=20T$ from the acceleration start ($T=2\pi/\omega$ laser period) . Frame~b): maximum ion energy vs time \cite{macchiPPCF13} (reproduced by permission of IOP Publishing). Both simulations have been performed for a $9\lambda\times(10\lambda)^2$ pulse (FWHM values) with peak amplitude $a_0=198$ and circular polarization, and a hydrogen plasma foil with surface density $\sigma=64m_pn_c\lambda$, so that $a_0\simeq\zeta$. See references for details.
}
\label{fig:3Dsim}
\end{figure}

Simulations by Tamburini et al. \cite{tamburiniPRE12} (Fig.\ref{fig:3Dsim}~a) have given a first evidence of the energy enhancement in fully 3D simulations. These simulations also indicated a baseline for LS operation in the relativistic regime, showing that the use of CP lead to higher energies and a more collimated beam with respect to the LP case, and also to negligible radiation friction effects. To evaluate the energy gain at the end of the acceleration stage, larger computational resources have been necessary to extend the simulation time by four times. In such simulations, the temporal dependence of maximum ion energy is in good agreement with the $\sim t^{3/5}$ scaling given by Eq.(\ref{eq:LSunlimitedscalings}), as shown in Fig.\ref{fig:3Dsim}~b) \cite{macchiPPCF13}. Ultimately, the acceleration is stopped by the onset of transparency. These simulations have been performed on the FERMI supercomputer at CINECA, Italy. 

\section{Conclusions and perspectives}

The laser-driven light sail concept, which was first studied as a visionary approach to interstellar travel, presently represents an implementation of Veksler's coherent acceleration paradigm and a possible route towards a laser-plasma accelerator. Experiments are just entering in the regime of intensities exceeding $5\times 10^{20}~\mbox{W cm}^{-2}$ where, according to our discussion, the radiation pressure push is expected to be the dominant acceleration mechanism. 
Recent progress in both achieving extremely high-contrast pulses and manufacturing ultrathin targets has been crucial for light sail experiments \cite{henigPRL09,dollarPRL12,karPRL12,aurandNJP13,steinkePRSTAB13} and results such as the observation of the fast scaling of ion energy in the non-relativistic limit \cite{karPRL12} are promising. However, several open issues are apparent, such as achieving monoenergetic spectra, and the effect of parameters such as the laser pulse focusing and duration still needs to be completely understood and optimized.

With the availability of next generation lasers at extreme intensities, success of the light sail approach in producing relativistic ions will depend on the possibility to achieve and control the so-called ``unlimited'' regime based on a suitable (and possibly self-regulated) transverse expansion of the target. On this route one expects technical challenges, such as clean circular polarization for ultraintense pulses, as well as other possible issues not considered in this paper such as the target stability.

\section*{Acknowledgments}
It is a pleasure to thank S. V. Bulanov and F. Pegoraro for scientific inspiration and enlightening discussions, and T. V. Liseykina, A. Sgattoni, S. Sinigardi and M. Tamburini for hard simulation work.
Support from the Italian Ministry of University and Research via
the FIR project ``SULDIS'' is acknowledged.

\end{document}